\documentclass[journal]{IEEEtran}

\usepackage{amsmath,amssymb,amsfonts}
\usepackage{algorithmic, algorithm}
\usepackage{cite, amsfonts, amssymb, amsthm, bm, bbm, graphicx, relsize, multirow, booktabs, tikz,soul}
\usepackage{subcaption}

\ifCLASSINFOpdf
\else
\fi

\theoremstyle{definition}

\theoremstyle{remark}

\newcommand{\ds}{\displaystyle}

\begin{document}

	\bstctlcite{IEEEexample:BSTcontrol}
	
	This paper was originally submitted for publication to the IEEE Wireless Communications Letters on April 14, 2020. After a rejection and resubmission, it was finally  accepted for publication on August 25, 2020.  
	
	\bigskip

	\copyright 2020 IEEE. Personal use of this material is permitted. Permission from IEEE must be obtained for all other uses, in any current or future media, including reprinting/republishing this material for advertising or promotional purposes, creating new collective works, for resale or redistribution to servers or lists, or reuse of any copyrighted  component of this work in other works.”

\title{Pilot Assignment in Cell-Free Massive MIMO based on the Hungarian Algorithm}

\author{Stefano Buzzi,~\IEEEmembership{ Senior Member,~IEEE,}
        Carmen D'Andrea,~\IEEEmembership{ Member,~IEEE,} Maria Fresia, 
      Yong-Ping Zhang,~\IEEEmembership{ Member,~IEEE,} and Shulan Feng
      \thanks{This work was supported by HiSilicon through cooperation agreement YBN2018115022.}
\thanks{S. Buzzi and C. D'Andrea are with the Department
of Electrical and Information Engineering, University of Cassino and Southern Latium, Cassino,
Italy, and with Consorzio Nazionale Interuniversitario per le Telecomunicazioni (CNIT), Parma, Italy.
M. Fresia is with Wireless Terminal Chipset Technology Lab Huawei Technol. Duesseldorf GmbH, Munich, Germany. Y. Zhang and S. Feng  are with  Balong Standard and Patent Department, HiSilicon Technologies, China.}}

\maketitle

\begin{abstract}
This letter focuses on the problem of pilot assignment in cell-free massive MIMO systems. Exploiting the well-known Hungarian algorithms, two procedures are proposed, one maximizing the system throughput, and the other one  maximizing the fairness across users.  {The algorithms operate based on the knowledge of large-scale fading coefficients as a proxy for the distances between users in the system, and take into account both the uplink and downlink performance.} Numerical results show that the proposed pilot assignment algorithms are effective and outperform the many competing alternatives available in the literature. 
\end{abstract}

\begin{IEEEkeywords}
cell-free, massive MIMO, pilot assignment, Hungarian algorithm, wireless networks, B5G, 6G
\end{IEEEkeywords}

%
\IEEEpeerreviewmaketitle

\section{Introduction}

Cell-free (CF) massive MIMO (mMIMO) is a wireless network deployment architecture credited to be a possible evolution of traditional multicell mMIMO systems \cite{Ngo_CellFree2016,BuzziWCL2017}. In  CF mMIMO, a very large number of distributed single-antenna access-points (APs) serves several mobile stations (MSs) using the same time-frequency resource. All APs are connected to a central processing unit (CPU) and cooperate via a backhaul network, and time-division duplex (TDD) protocol is used. CF mMIMO systems have actually no  cell boundaries and  benefit from large-scale fading (LSF) diversity. They are thus able to ensure an improved level of fairness across users when compared with multicell mMIMO systems\cite{BuzziWCL2017,bjornson2019CF_MMSE,Buzzi_TWC2019}. 

Similarly to multicell mMIMO, the performance of CF mMIMO systems is critically affected by the lack of a sufficiently large number of orthogonal pilot sequences, which prevents the possibility of acquiring channel state information (CSI) with no interference. The use of properly designed pilot assignment (PA) algorithms, thus, is crucial in order to ensure good performance in highly loaded networks. The problem of PA in CF mMIMO was firstly investigated in \cite{Ngo_CellFree2016}, where, starting from a random PA (RPA), the authors propose a greedy pilot assignment based on the knowledge of the LSF channel coefficients that iteratively updates the pilot of the worst performing MS in order to increase the system fairness. The authors of  \cite{LB_Greedy_ZhangICC2018}, instead,  propose to use the algorithm in \cite{Ngo_CellFree2016} using as starting point an assignment based on the location of the MSs. Similarly, patent \cite{Ngo_marzetta_patent2017} proposed an iterative algorithm, based on consecutive updates of the pilots for the worst and best performing MSs, again aiming at the maximization of the system fairness. 
In \cite{attarifar2018random} a PA algorithm based on the knowledge of the MSs' positions is proposed  {and in \cite{Femenias_Access2019} a similar procedure is considered taking into account the LSF coefficients of the channels between MSs and APs. }

The Hungarian algorithm \cite{munkres1957algorithms}, a popular combinatorial algorithm used to solve weighted matching problems in a bipartite graph with polynomial complexity, has been  used to solve the PA problem in traditional multicell mMIMO systems \cite{Nguyen_Access2015,Ma_PilotAss_massiveMIMO2018}; in these papers, however the number of orthogonal pilots in each cell is assumed to be greater than the number of users, i.e., there is no intra-cell pilot contamination.  

{The contribution of this paper is the development of a new PA iterative procedure involving at each step the definition of a proper bipartite graph such that the Hungarian algorithm can be used to perform matching. Two new expressions for the reward coefficients matrix are introduced, capable of jointly taking into account the system performance on the uplink (UL) and on the downlink (DL), and maximizing the system throughput and the system fairness, respectively.  The proposed  algorithms are useful when the number of orthogonal pilot sequences is significantly lower than the number of active users in the area, i.e., there is strong pilot contamination.} The proposed strategies exploit the knowledge of the LSF coefficients between the APs and the MSs as a proxy of the distances between the MSs.

The numerical results, provided in Section IV, will reveal the superiority of the newly proposed solutions with respect to competing alternatives.

\section{System model and performance measures}
We consider an area with $K$ single-antenna MSs and $M$ APs each with $N_{\rm AP}$ antennas connected, by means of a backhaul network, to a CPU wherein data-decoding is performed. We denote by $\mathcal{K}_m$ and $\mathcal{M}_k$ the set of MSs served by the $m$-th AP, and the set of APs serving the $k$-th MS, respectively.
The symbol $\mathbf{g}_{k,m}$  denotes the $N_{\rm AP}$-dimensional vector representing the channel between the $k$-th MS and the $m$-th AP; we assume $\mathbf{g}_{k,m}=\sqrt{\beta_{k,m}} \mathbf{h}_{k,m}$, with $\mathbf{h}_{k,m}$ an $N_{\rm AP}$-dimensional vector whose entries are i.i.d ${\cal CN}(0,1)$ random variables (RVs), modeling the fast fading, and $\beta_{k,m}$ the LSF coefficient.

 {At each AP, a channel estimate of the channel $\mathbf{g}_{k,m}$, say $\widehat{\mathbf{g}}_{k,m}$, is obtained through a linear minimum-mean-square-error (MMSE) processing as reported in \cite{Ngo_CellFree2016,BuzziWCL2017}. Denote by $\tau_p< \tau_c $ the length (in time-frequency samples) of the UL training phase and by $\tau_c$ the length (in time-frequency samples) of the coherence interval, and assume that the pilot sequences transmitted by the MSs are chosen in a set of $\tau_p$ orthogonal sequences $\mathcal{P}_{\tau_p}=\left\lbrace \boldsymbol{\varphi}_1, \boldsymbol{\varphi}_2, \ldots, \boldsymbol{\varphi}_{\tau_p} \right\rbrace$, where $\boldsymbol{\varphi}_i$ is the $i$-th unit norm $\tau_p$-dimensional pilot  sequence.}

On the DL, the APs treat the channel estimates as the true channels and perform conjugate beamforming, while 
on the UL, the generic $m$-th AP participates to the decoding of the data sent by the MSs in ${\cal K}_m$, but data decoding takes place in the CPU \cite{BuzziWCL2017,Buzzi_TWC2019}.

\begin{figure*}
\begin{equation}
\mathcal{R}_{k}^{\rm DL}= \ds\frac{\tau_{d}}{\tau_c}  W \log_2 \left( 1+ \frac{  \left( \ds \sum_{m\in{\cal M}_k} {\ds  \eta_{k,m}^{\rm DL} \gamma_{k,m}} \right)^2}{
\ds \sum_{j=1}^K \sum_{m\in{\cal M}_j}  \eta_{j,m}^{\rm DL}  \beta_{k,m} \gamma_{j,m} + \ds \sum_{\substack{j=1 \\ j\neq k}}^K \left( \ds \sum_{m \in {\cal M}_j} \eta_{j,m}^{\rm DL} \sqrt{ \frac{\eta_k}{\eta_j}} \gamma_{j,m} \frac{\beta_{k,m}}{\beta_{j,m}}\right)^2 \left|\boldsymbol{\varphi}_k^H \boldsymbol{\varphi}_j\right|^2 + \sigma^2_z }
 \right) 
\label{eq:SE_DL}
\end{equation}
\begin{equation}
\mathcal{R}_{k}^{\rm UL}= \ds\frac{\tau_{u}}{\tau_c}  W \log_2 \left( 1+ \frac{ \eta_{k}^{\rm UL} \left( \ds \sum_{m\in{\cal M}_k} {\ds  \gamma_{k,m}} \right)^2}{
\ds \sum_{j=1}^K \eta_{j}^{\rm UL}  \sum_{m\in{\cal M}_k} \beta_{j,m} \gamma_{k,m} + \ds \sum_{\substack{j=1 \\ j\neq k}}^K\eta_{j}^{\rm UL}\left( \ds \sum_{m \in {\cal M}_k}  \gamma_{k,m} \frac{\beta_{j,m}}{\beta_{k,m}}\right)^2 \left|\boldsymbol{\varphi}_j^H \boldsymbol{\varphi}_k\right|^2 + \sigma^2_w \!\!\!\!\sum_{m\in{\cal M}_k} {\!\! \gamma_{k,m}}
} \right) 
\label{eq:SE_UL}
\end{equation}
\hrulefill
\end{figure*}

As performance measures used for the testing of the proposed PA algorithms we will consider the achievable rates in DL and UL. Applying the use-and-then-forget (UatF) bounding techniques in \cite{marzetta2016fundamentals}, a lower-bound to the $k$-th MS DL and UL achievable rates is reported in Eqs. \eqref{eq:SE_DL} and \eqref{eq:SE_UL} at the top of the next page, respectively. In these expressions, the following notation has been used: $W$ is the system bandwidth, $\tau_{d}$ and $\tau_{u}$ are the lengths (in time-frequency samples) of the DL and UL data transmission phases in each coherence interval;  $\eta_{k,m}^{\rm DL}$ a scalar coefficient controlling the power transmitted by the $m$-th AP to the $k$-th MS; $\sigma^2_z$ is the AWGN noise variance at the generic MS receiver; ${\eta_{k}^{\rm UL}}$ is the UL transmit power used by the $k$-th MS in the data transmission phase; $\sigma^2_w$ is the AWGN noise variance at the generic AP receiver; ${\eta}_k$ is the power employed by the $k$-th MS during the training phase, $\boldsymbol{\varphi}_k$ is the $\tau_p$-dimensional column pilot sequence transmitted by the $k$-th MS and $
\gamma_{k,m}=\mathbb{E}\left[ \widehat{\mathbf{g}}_{k,m}^H \widehat{\mathbf{g}}_{k,m}\right]$. Details on the UatF bound and on the derivations of Eqs. \eqref{eq:SE_DL} and \eqref{eq:SE_UL} can be found in\cite{marzetta2016fundamentals,Ngo_CellFree2016,Buzzi_TWC2019} and are here omitted due to the lack of space.

\section{Pilot assignment algorithm}

We are now ready to illustrate the proposed PA procedures. The schemes that we propose are iterative, have a common structure, and start with a random PA. 
Basically, the steps of the algorithms can be stated as follows:
\begin{enumerate}
\item Assign to each MS a pilot randomly picked from the set $\mathcal{P}_{\tau_p}$ of orthogonal pilots.
\item Consider the generic $k$-th MS; pick the $\tau_p-1$ MSs that are \textit{closest} to MS $k$. The set of these MSs, including the $k$-th one,  forms the set $\mathcal{S}_k$, of cardinality $\tau_p$. The remaining $K-\tau_p$ MSs are grouped in the set $\mathcal{T}_k$.
\item Assign pilots to the users in the set $\mathcal{S}_k$ considering the PA of the users in the set $\mathcal{T}_k$ as fixed. 
\item Repeat steps 2) and 3) for all values of $k=1, \ldots, K$.
\item Repeat steps from 2) to 4) until the performance measures have reached convergence and/or the maximum number of allowed iterations has been reached. 
\end{enumerate}
We now provide further details to better clarify the meaning of the above steps.

\subsection{Defining the set $\mathcal{S}_k$} 
To execute the above step 2), the $(\tau_p-1)$ MSs that are \textit{closest} to the $k$-th MS are to be selected. 
One simple way of doing this is to rely on the knowledge of the MSs' positions. Indeed, if this knowledge is available at the CPU, the set $\mathcal{S}_k$ can be readily defined. 

If, instead, we want to avoid relying on MSs' location information, knowledge of the LSF coefficients can be exploited as indicators of the distance between MSs. Precisely, we are not able to select the $(\tau_p -1)$ MSs that are closest to the $k$-th MS, but only the $(\tau_p-1)$ MSs that are closest to (i.e., have the largest LSF coefficients to) the AP that is closest to MS $k$. The two sets of course cannot be claimed to be coincident but with high likelihood will have several common elements. 
 {Extensive numerical experiments, not reported here for the sake of brevity, have confirmed that using LSF coefficients instead of true MSs positions leads to a practically imperceptible performance loss, and this is why in this paper we just focus on the exposition of the algorithms exploiting the LSF coefficients knowledge.}
More precisely, the procedure works as follows. For the $k$-th MS, the CPU first computes the index of its nearest AP as $m^*=\arg \max_{m} \; \; \beta_{k,m}$.
Then, consider the set of the LSF coefficients $\mathcal{D}_{\overline{k},m^*}=\left\{\beta_{j,m^*}\right\}_{j=1, j \neq k}^K$, whose entries are sorted in decreasing order, and denote by $U_{m^*,k}(\ell)$ the MS index associated with the LSF coefficient appearing in the $\ell$-th position of the set $\mathcal{D}_{\overline{k},m^*}$. The set $\mathcal{S}_k$ will thus contain the index (MS) $k$ and the indexes (MSs) associated to the $(\tau_p-1)$ largest coefficients in $\mathcal{D}_{\overline{k},m^*}$, i.e.: $\mathcal{S}_k= \left\lbrace k, U_{m^*,k}(1), U_{m^*,k}(2), \ldots, U_{m^*,k}(\tau_p-1) \right\rbrace$.

\subsection{Running the Hungarian algorithm} \label{Hungarian_explanation}
Once the sets $\mathcal{S}_k$ and $\mathcal{T}_k$ have been defined, the set of $\tau_p$ available orthogonal pilots is to be assigned to the $\tau_p$ MSs in $\mathcal{S}_k$ according to some optimality criterion. 
Denoting by $a_{\ell,q}^{(k)}$ a scalar quantity measuring the \textit{reward} (to be specified in the following subsection) for the system if the $q$-th pilot 
in $\mathcal{P}_{\tau_p}$ is assigned to the $\ell$-th MS in the set $\mathcal{S}_k$, 
and letting ${x_{\ell,q}^{(k)}}$ be a binary $0-1$ variable indicating that the $q$-th pilot sequence is assigned to the $\ell$-th MS,
we are formally faced with the following 
optimization problem:
\begin{subequations}\label{Prob:Matching_problem}
\begin{align}
\ds\max_{x_{\ell,q}^{(k)} \in \{0,1\}}\;&\ds \sum_{\ell=1}^{\tau_p}\sum_{q=1}^{\tau_p} x_{\ell,q}^{(k)} a_{\ell,q}^{(k)} \label{Prob:aMatching_problem}\\
\;\textrm{s.t.}\;& \sum_{\ell=1}^{\tau_p} x_{\ell,q}^{(k)} =1 \; \forall \, q,
\; \text{and} \; \sum_{q=1}^{\tau_p} x_{\ell,q}^{(k)} =1 \; \forall \, \ell . \label{Prob:bMatching_problem}
\end{align}
\end{subequations}
Problem \eqref{Prob:Matching_problem} accepts as an input the coefficients 
$a_{\ell,q}^{(k)}$, for all $\ell$ and $q$, and solving it entails providing the values of the optmization variables $x_{\ell,q}^{(k)}$, for all $\ell$ and $q$.

The constraints in \eqref{Prob:bMatching_problem} are needed to ensure that each pilot is assigned to just one user and that all the pilots for the MSs in $\mathcal{S}_k$  are used once, respectively. 
One way to solve the above combinatorial optimization problem in polynomial time is to resort to the Hungarian method \cite[Algorithm 14.2.3]{jungnickel2007graphs}. A fast and efficient implementation of the Hungarian algorithm was introduced in \cite{munkres1957algorithms}. We do not provide further details on this algorithm for the sake of brevity. 

\subsection{Defining the reward coefficients}
Let us now define how the coefficients $a_{\ell,q}^{(k)}$ are computed.
 {If the goal is to maximize the system throughput, then a reasonable choice is to assume that $a_{\ell,q}^{(k)}$ is equal to the product between the $\ell$-th MS DL and UL rates when it is assigned the $q$-th pilot, i.e. we have
$
a_{\ell,q}^{(k)}=\ds  \mathcal{R}_{\ell}^{\rm DL}(\{ x_{\ell,q}=1\}) \mathcal{R}_{\ell}^{\rm UL}(\{ x_{\ell,q}=1\})
$,
where $\mathcal{R}_{\ell}^{\rm x}(\{ x_{\ell,q}=1\})$ denotes the $\ell$-th MS rate when it is assigned the $q$-th pilot, where  ${\rm x}$ can be ${\rm DL}$ or ${\rm UL}$.}
Note that the above quantity does not depend on the assignments that have been done for the other MSs in 
 $\mathcal{S}_k$, since these MSs are using orthogonal pilots; rather, it will depend on the locations of the MSs in $\mathcal{T}_k$ that are assigned the same $q$-th pilot as the MS $k$. We refer to this PA strategy as sum-rate maximizing Hungarian PA (SHPA).

 {If, instead, the system designer goal is to maximize fairness across users, a different strategy is in order. Denote by $\widetilde{\mathcal{T}}_k(q)$ the set of MSs in $\mathcal{T}_k$ that are using the $q$-th pilot, and let
$
a_{\ell,q}^{(k)}= \min_{j \in \widetilde{\mathcal{T}}_k(q) \cup \{\ell\}} \mathcal{R}_j^{\rm DL} (\{ x_{\ell,q}=1\}) \mathcal{R}_j^{\rm UL} (\{ x_{\ell,q}=1\}).
$}
Otherwise stated, $a_{\ell,q}^{(k)}$ is the smallest product between the DL and UL rates computed among all the MSs in the system that are using the $q$-th pilot, including the $\ell$-th MS. We refer to this PA strategy as minimum-rate maximizing Hungarian PA (MHPA).
\begin{table}[t!]
\centering
\caption{ {Average number of iterations needed to reach convergence.}}
\label{table:num_iterations}
\begin{tabular}{|l|l|l|}
\hline
                        & \textbf{SHPA} & \textbf{MHPA} \\ \hline
$K=20$    &        3.9                   &        4.7               \\ \hline
$K=40$   &          4.1                 &          6.6             \\ \hline
$K=60$     &        4.1                  &          8.1             \\ \hline
\end{tabular}
\end{table}

\section{Numerical Results}
In our simulation setup, we assume a communication bandwidth $W = 20$ MHz centered over the carrier frequency $f_0=1.9$ GHz. The antenna height at the AP is $10$ m and at the MS is $1.65$ m. The additive thermal noise is assumed to have a power spectral density of $-174$ dBm/Hz, while the front-end receiver at the APs and at the MSs is assumed to have a noise figure of $9$ dB. 
We consider $M=100$, $N_{\rm AP}=4$ $K=40$ and a MS-centric approach \cite{BuzziWCL2017,Buzzi_TWC2019}, where each MS is served by the $N=20$ APs with the highest LSF coefficients and $\mathcal{K}_m$ and $\mathcal{M}_k$ are defined accordingly. 
The APs and MSs are deployed at random positions on a square area of $1000 \times 1000$ (square meters). In order to avoid boundary effects, the square area is wrapped around \cite{Ngo_CellFree2016,BuzziWCL2017}. The LSF coefficient $\beta_{k,m}$ is modelled as in \cite[Table B.1.2.2.1-1]{3GPP_36814_GUE_model} and the shadow fading coefficients from an AP to different MSs are correlated as in \cite[Table B.1.2.2.1-4]{3GPP_36814_GUE_model}. The shadow fading correlation among MSs is instead modeled as in \cite{bjornson2019CF_MMSE}. The orthogonal pilot sequences in $\mathcal{P}_{\tau_p}$ have length $\tau_p=8$; the DL and UL data transmission phases durations in samples are $\tau_{u}=\tau_{d}=\frac{\tau_c-\tau_p}{2}$, with $\tau_c=200$. The UL transmit power for channel estimation is $\eta_k=\tau_p p_k$, with $p_k=100$ mW, $\forall k=1,\ldots,K$.  {Regarding power control, on the DL,
the power coefficients are set as  $\eta_{k,m}^{\rm DL}=\gamma_{k,m}P_{m}^{\rm DL}/(\sum_{k \in \mathcal{K}(m)} \gamma_{k,m})$ for the case in which PA aims at the sum-rate maximization, and as $\eta_{k,m}^{\rm DL}=\gamma_{k,m}^{-\left(\alpha_{\rm DL}+1\right)}P_{m}^{\rm DL}/(\sum_{k \in \mathcal{K}(m)} \gamma_{k,m}^{-\alpha_{\rm DL}})$, with $\alpha_{\rm DL}=-0.5$, when the PA aims at minimum-rate maximization. We also let
$P_{m}^{\rm DL}=200$ mW, $\forall \; m=1,\ldots,M$.
For the UL, instead, we let $\eta_{k}^{\rm UL}=\min\left(P_{\rm max}^{\rm UL}, P_0 \bar{\gamma}_k^{- \alpha_{\rm UL}}\right)$, $\forall \; k=1,\ldots,K$, with $P_0=-10$ dBm, $\alpha_{\rm DL}=0.5$, $\bar{\gamma}_k=\sqrt{\sum_{m \in \mathcal{M}_k} \gamma_{k,m}}$ and $P_{\rm max}^{\rm UL}=100$ mW.}

 {
First of all, we evaluate the algorithms complexity. Steps 2) and 3) detailed at the beginning of Section III require $K$ runs of the Hungarian algorithm, each one having  complexity  $O\left(\tau_p^3\right)$\cite[Algorithm 14.2.3]{jungnickel2007graphs}. The complexity of the steps from 2) to 4) is thus   $O\left(K \tau_p^3\right)$. Table \ref{table:num_iterations} reports the number of times that steps from 2) to 4) (i.e., the  number of iterations) are needed in order to reach convergence for different number of users $K$. It is seen that 
the number of iterations is weakly dependent on $K$; we can thus state that the proposed algorithms complexity is approximately proportional to $K\tau_p^3$.} 

Next, we compare the performance of the proposed PA algorithms with a RPA and with the solutions in\cite{Ngo_CellFree2016, LB_Greedy_ZhangICC2018,Ngo_marzetta_patent2017,attarifar2018random, Femenias_Access2019}. The reward coefficients are defined using the rate expressions reported in  Eqs. \eqref{eq:SE_DL} and  \eqref{eq:SE_UL}.

\begin{figure}[t!]
\begin{subfigure}{0.5\textwidth}
  \centering
  \includegraphics[scale=0.42]{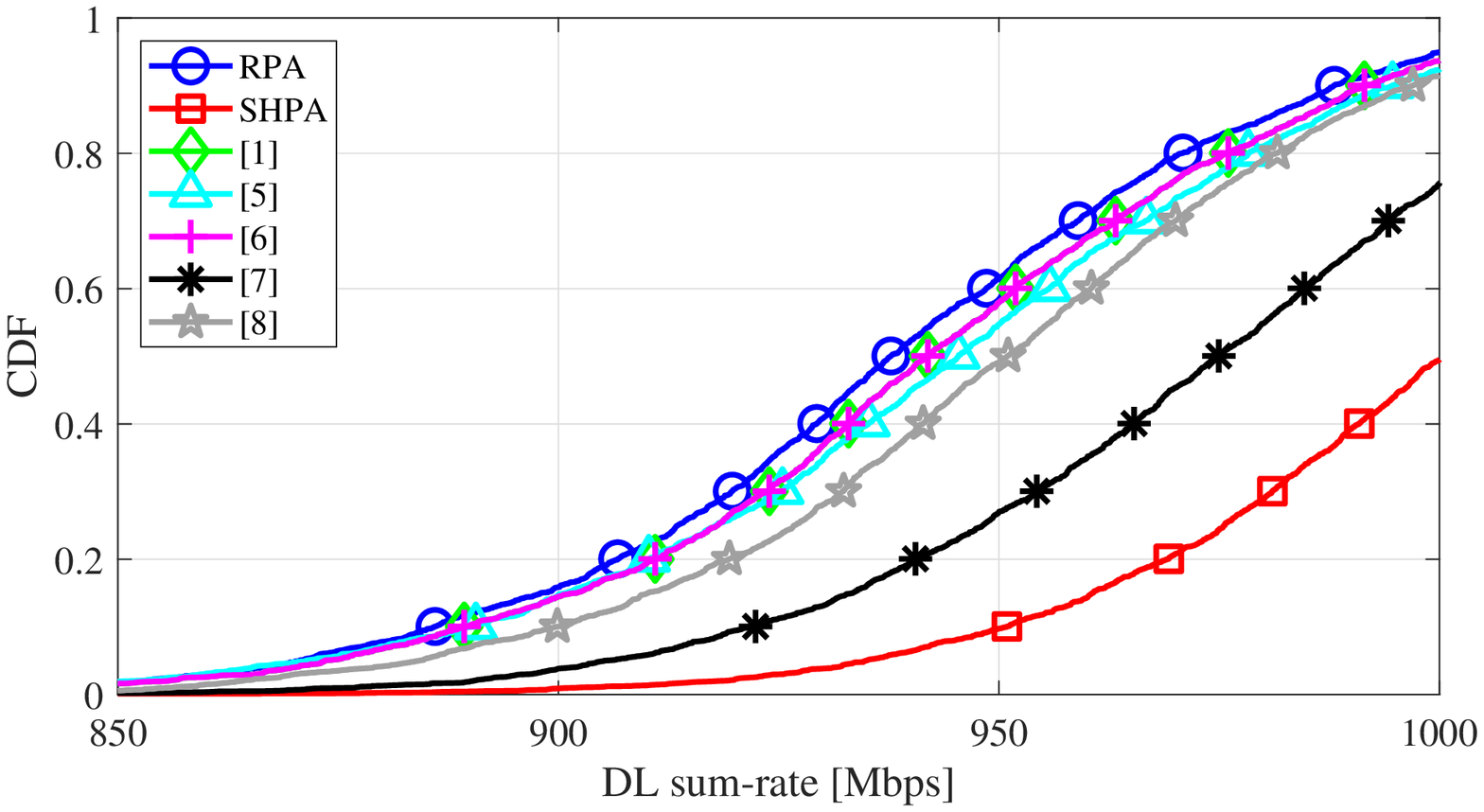}  
  \caption{\textcolor{white}{.}}
\end{subfigure}
\begin{subfigure}{0.5\textwidth}
  \centering
  \includegraphics[scale=0.42]{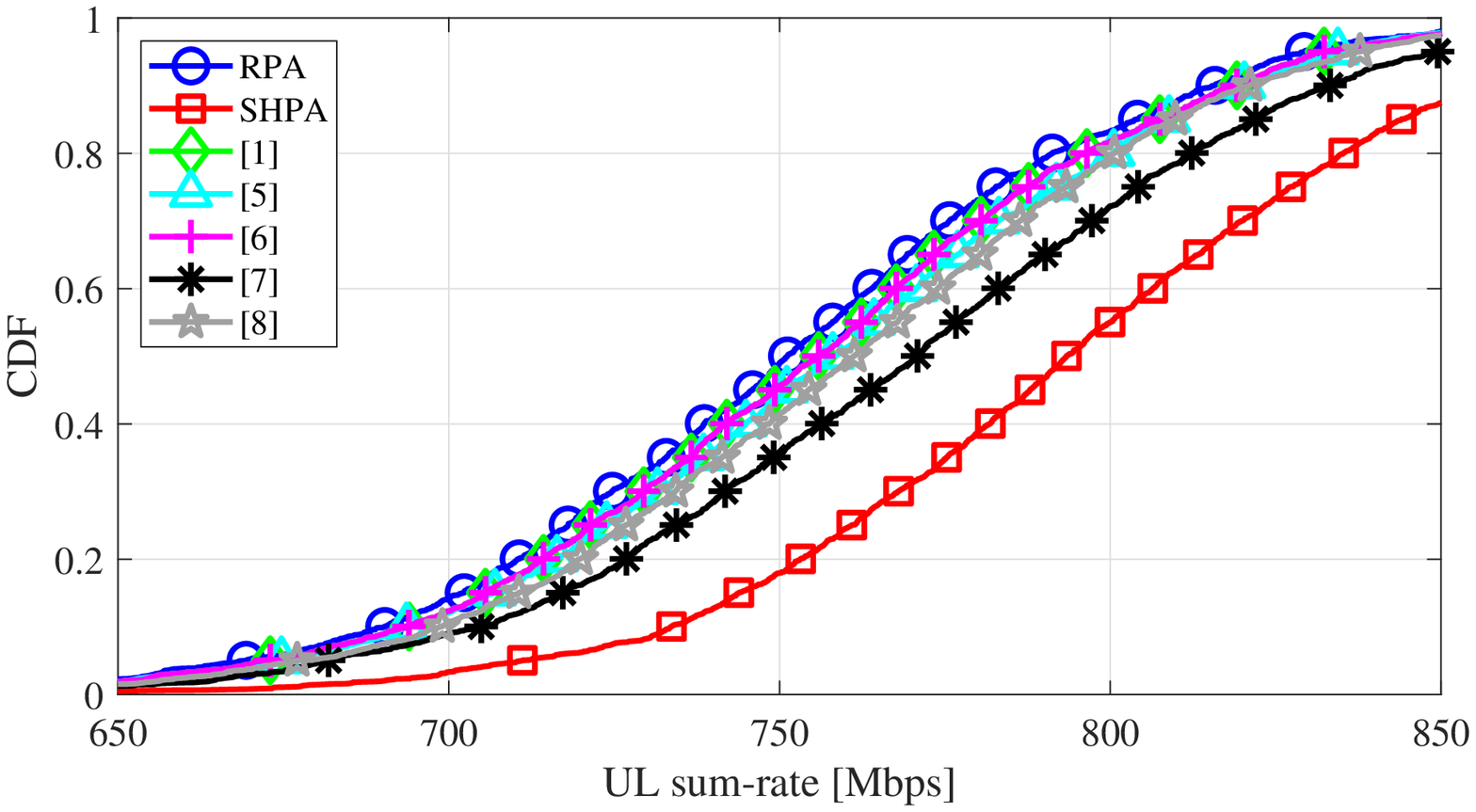}  
  \caption{\textcolor{white}{.}}
\end{subfigure}
\caption{CDFs of DL and UL sum-rate. Parameters: $M=100$, $N_{\rm AP}=4$, $K=40$, $\tau_p=8$.}
\label{Fig:sum_rate}
\end{figure}

\begin{figure}[t!]
\begin{subfigure}{.5\textwidth}
  \centering
  \includegraphics[scale=0.42]{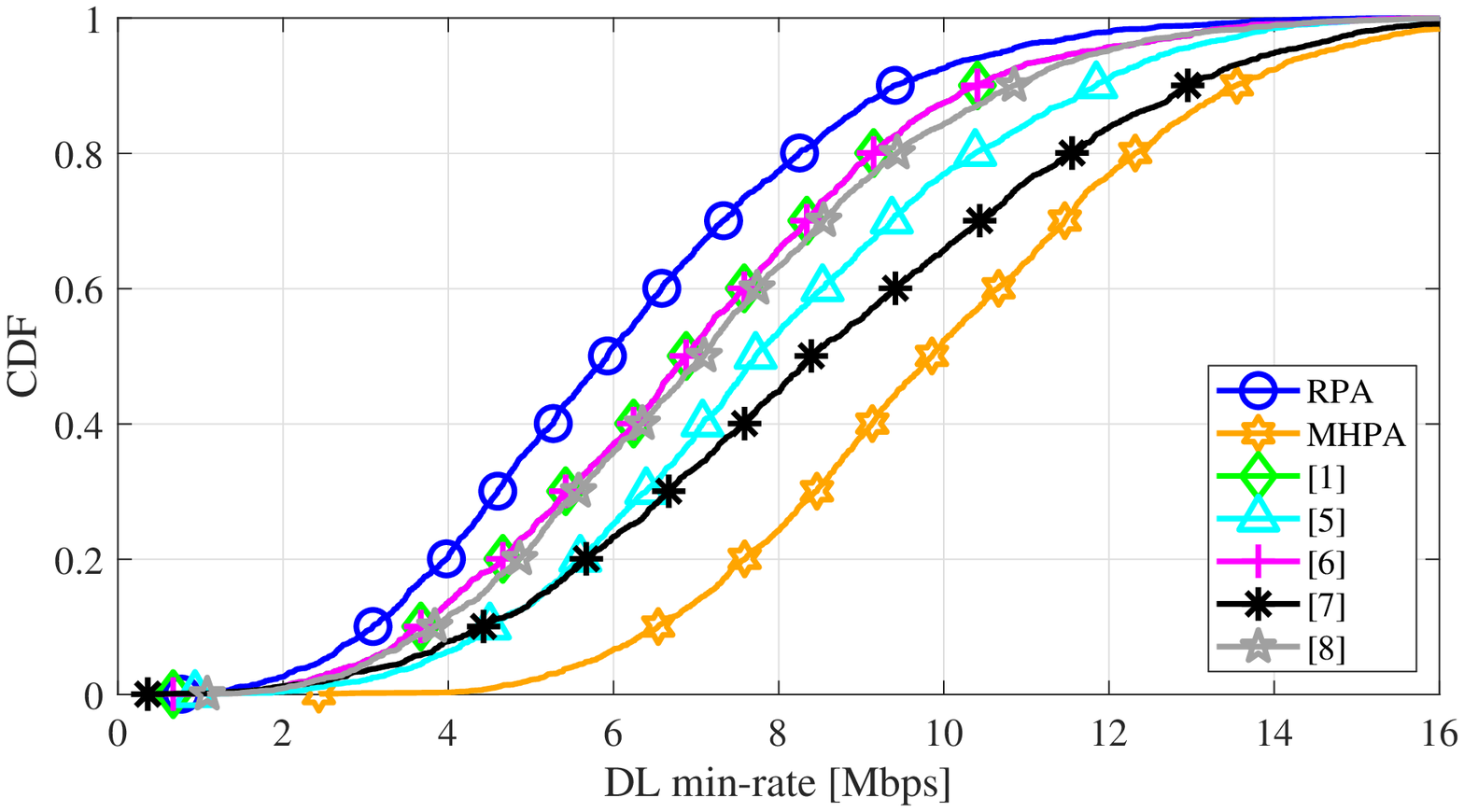}  
  \caption{\textcolor{white}{.}}
\end{subfigure}
\begin{subfigure}{.5\textwidth}
  \centering
  \includegraphics[scale=0.42]{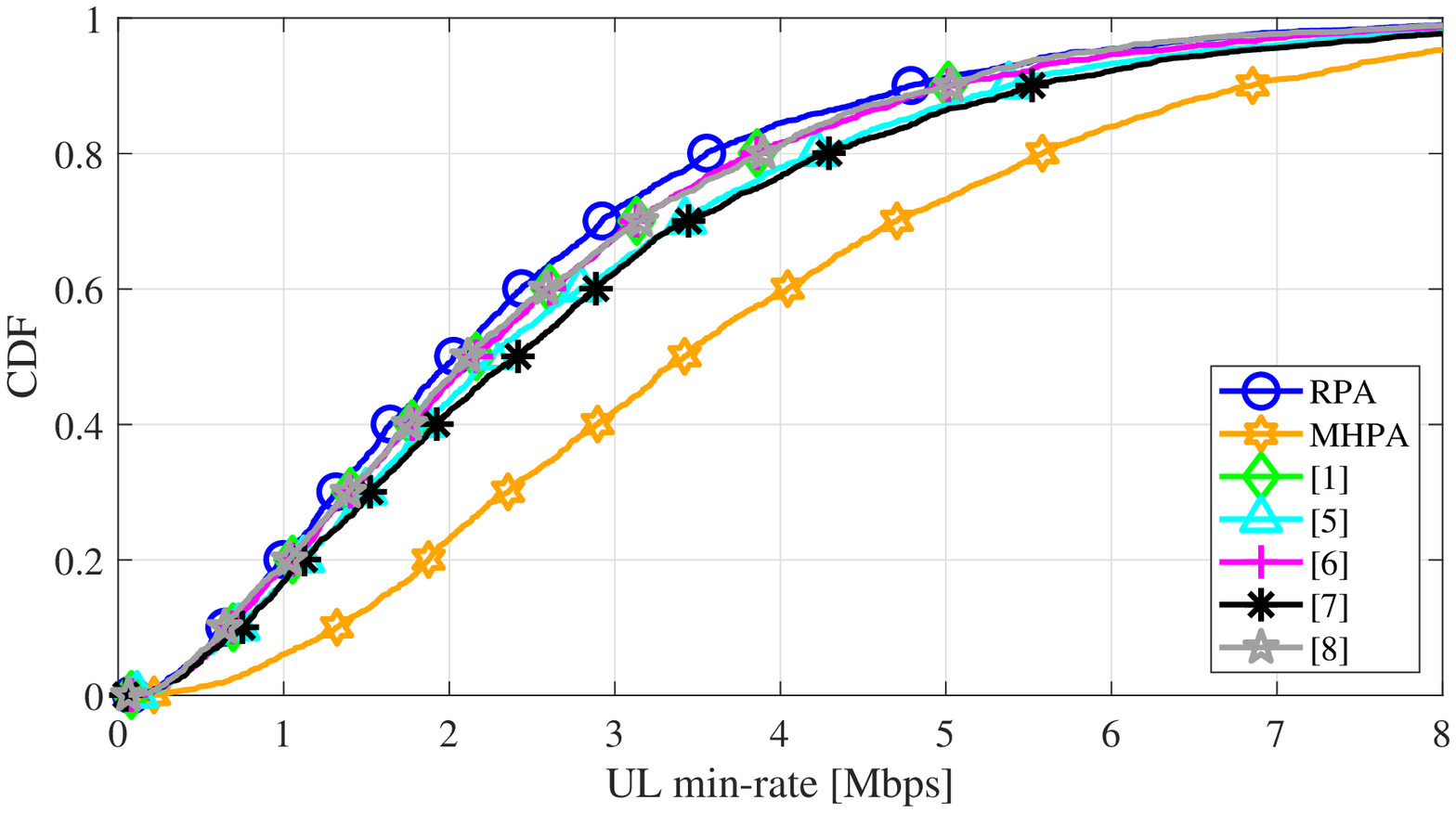}  
  \caption{\textcolor{white}{.}}
\end{subfigure}
\caption{CDFs of DL and UL min-rate. Parameters: $M=100$, $N_{\rm AP}=4$, $K=40$, $\tau_p=8$.}
\label{Fig:min_rate}
\end{figure}

Figs. \ref{Fig:sum_rate} and \ref{Fig:min_rate} report the sum-rate and min-rate cumulative distribution functions (CDFs) for the DL and UL. Inspecting the figures we can see that the proposed solutions outperform competing alternatives both in terms of sum-rate and min-rate. 
To have an insight into the values of the per-user rates, 
in Table \ref{table:5_percent_rate} we report the 5\%-rate performance obtained with the different PA strategies. It is seen that there are improvements in the order of 15$\%$ with respect to existing competing alternative.

\begin{table}[t]
\centering
\caption{DL and UL 5\%-rate. Parameters: $M=100$, $N_{\rm AP}=4$, $K=40$, $\tau_p=8$.}
\label{table:5_percent_rate}
\begin{tabular}{|l|l|l|l|l|}
\hline
                        & \textbf{DL 5\%-rate (MR/SR Max)} & \textbf{UL 5\%-rate (MR/SR Max)} \\ \hline
RPA    &        9 Mbps / 6.3 Mbps                  &        3.9 Mbps  / 3.9 Mbps           \\ \hline
Proposed     &        12  Mbps / 9.4 Mbps                   &          5.1 Mbps / 5 Mbps        \\ \hline
[1]    &           9.5 Mbps / 7 Mbps             &            4 Mbps  / 4 Mbps           \\ \hline
[5]  &           10.2 Mbps  / 7.4 Mbps                 &             4.1 Mbps  / 4.1 Mbps          \\ \hline
[6] &            9.5 Mbps   / 7 Mbps               &                4 Mbps/ 4 Mbps         \\ \hline
[7]  &           10.4 Mbps  / 8.4 Mbps                 &             4.2 Mbps / 4.2 Mbps           \\ \hline
[8] &            10.3 Mbps   / 7 Mbps               &                3.9 Mbps  / 3.9 Mbps       \\ \hline
\end{tabular}
\end{table}

\vspace{-0.5cm}
\section{Conclusion}
In this letter, the problem of PA in a CF mMIMO system has been considered. An iterative procedure based on the Hungarian algorithm has been proposed. The algorithm parameters can be tuned so as to maximize either the sum-rate or the fairness across users, and can be implemented based on the knowledge of the LSF coefficients. Simulation results have shown that the proposed procedures exhibit a significant advantage over several competing alternatives.


\bibliography{References}

\begin{thebibliography}{10}
\providecommand{\url}[1]{#1}
\csname url@samestyle\endcsname
\providecommand{\newblock}{\relax}
\providecommand{\bibinfo}[2]{#2}
\providecommand{\BIBentrySTDinterwordspacing}{\spaceskip=0pt\relax}
\providecommand{\BIBentryALTinterwordstretchfactor}{4}
\providecommand{\BIBentryALTinterwordspacing}{\spaceskip=\fontdimen2\font plus
\BIBentryALTinterwordstretchfactor\fontdimen3\font minus
  \fontdimen4\font\relax}
\providecommand{\BIBforeignlanguage}[2]{{%
\expandafter\ifx\csname l@#1\endcsname\relax
\typeout{** WARNING: IEEEtran.bst: No hyphenation pattern has been}%
\typeout{** loaded for the language `#1'. Using the pattern for}%
\typeout{** the default language instead.}%
\else
\language=\csname l@#1\endcsname
\fi
#2}}
\providecommand{\BIBdecl}{\relax}
\BIBdecl

\bibitem{Ngo_CellFree2016}
H.~Q. Ngo, A.~Ashikhmin, H.~Yang, E.~G. Larsson, and T.~L. Marzetta,
  ``Cell-free massive {MIMO} versus small cells,'' \emph{IEEE Transactions on
  Wireless Communications}, vol.~16, no.~3, pp. 1834--1850, Jan. 2017.

\bibitem{BuzziWCL2017}
S.~Buzzi and C.~D'Andrea, ``Cell-free massive {MIMO}: User-centric approach,''
  \emph{IEEE Wireless Communications Letters}, vol.~6, no.~6, pp. 706--709,
  Dec. 2017.

\bibitem{bjornson2019CF_MMSE}
E.~Bj{\"o}rnson and L.~Sanguinetti, ``Making cell-free massive {MIMO}
  competitive with {MMSE} processing and centralized implementation,''
  \emph{IEEE Transactions on Wireless Communications}, vol.~19, no.~1, pp.
  77--90, Jan. 2020.

\bibitem{Buzzi_TWC2019}
S.~{Buzzi}, C.~{D'Andrea}, A.~{Zappone}, and C.~{D'Elia}, ``User-centric {5G}
  cellular networks: Resource allocation and comparison with the cell-free
  massive {MIMO} approach,'' \emph{IEEE Transactions on Wireless
  Communications}, vol.~19, no.~2, pp. 1250--1264, Feb. 2020.

\bibitem{LB_Greedy_ZhangICC2018}
Y.~{Zhang}, H.~{Cao}, P.~{Zhong}, C.~{Qi}, and L.~{Yang}, ``Location-based
  greedy pilot assignment for cell-free massive {MIMO} systems,'' in \emph{2018
  IEEE 4th International Conference on Computer and Communications (ICCC)},
  Dec. 2018, pp. 392--396.

\bibitem{Ngo_marzetta_patent2017}
A.~Ashikhmin, H.~Q. Ngo, T.~L. Marzetta, and H.~Yang, ``Pilot assignment in
  cell free massive {MIMO} wireless systems,'' Apr. 2017, {US} Patent
  9,615,384.

\bibitem{attarifar2018random}
M.~Attarifar, A.~Abbasfar, and A.~Lozano, ``Random vs structured pilot
  assignment in cell-free massive {MIMO} wireless networks,'' in \emph{2018
  IEEE International Conference on Communications Workshops (ICC
  Workshops)}.\hskip 1em plus 0.5em minus 0.4em\relax IEEE, 2018, pp. 1--6.

\bibitem{Femenias_Access2019}
G.~{Femenias} and F.~{Riera-Palou}, ``Cell-free millimeter-wave massive {MIMO}
  systems with limited fronthaul capacity,'' \emph{IEEE Access}, vol.~7, pp.
  44\,596--44\,612, Apr. 2019.

\bibitem{munkres1957algorithms}
J.~Munkres, ``Algorithms for the assignment and transportation problems,''
  \emph{Journal of the society for industrial and applied mathematics}, vol.~5,
  no.~1, pp. 32--38, 1957.

\bibitem{Nguyen_Access2015}
T.~M. {Nguyen}, V.~N. {Ha}, and L.~{Bao Le}, ``Resource allocation optimization
  in multi-user multi-cell massive {MIMO} networks considering pilot
  contamination,'' \emph{IEEE Access}, vol.~3, pp. 1272--1287, Aug. 2015.

\bibitem{Ma_PilotAss_massiveMIMO2018}
S.~{Ma}, E.~L. {Xu}, A.~{Salimi}, and S.~{Cui}, ``A novel pilot assignment
  scheme in massive {MIMO} networks,'' \emph{IEEE Wireless Communications
  Letters}, vol.~7, no.~2, pp. 262--265, Apr. 2018.

\bibitem{marzetta2016fundamentals}
T.~L. Marzetta, E.~G. Larsson, H.~Yang, and H.~Q. Ngo, \emph{Fundamentals of
  massive {MIMO}}.\hskip 1em plus 0.5em minus 0.4em\relax Cambridge University
  Press, 2016.

\bibitem{jungnickel2007graphs}
D.~Jungnickel, \emph{Graphs, networks and algorithms}.\hskip 1em plus 0.5em
  minus 0.4em\relax Springer Science \& Business Media, 2007, vol.~5.

\bibitem{3GPP_36814_GUE_model}
3GPP, ``Further advancements for {E-UTRA} physical layer aspects ({R}elease
  9),'' 3GPP TS 36.814, Tech. Rep., Mar. 2017.

\end{thebibliography}
\bibliographystyle{IEEEtran}

\end{document}